\documentclass[review]{elsarticle}

\usepackage{amssymb}
 \usepackage{amsthm}
\usepackage{tabularx}
\usepackage{hyperref}
\usepackage{graphicx}
\usepackage{subfigure}
\usepackage{booktabs} 
\usepackage[hang, small,labelfont=bf,up,textfont=it,up]{caption}
\usepackage{rotating}
\usepackage{relsize}
\usepackage{setspace}
\usepackage[a4paper,top=2.5cm,bottom=3cm,left=3cm,right=3cm]{geometry}
\usepackage{color}
\usepackage{soul}
\journal{Materials \& Design}

\begin{document}

\begin{frontmatter}
\ead{marco.beghi@polimi.it}
%\setlength{\textwidth}{16 cm}
%\journal{Material&Design}
%\begin{frontmatter}
\title{Amorphous, ultra-nano- and nano-crystalline tungsten-based coatings grown by Pulsed Laser Deposition: mechanical characterization by Surface Brillouin Spectroscopy}
\author{E. Besozzi$^a$, D. Dellasega$^{a,b}$, A. Pezzoli$^a$, C. Conti$^c$, M. Passoni$^{a,b}$, M.G. Beghi$^a$}

%----
\address{$^a$ Dipartimento di Energia, Politecnico di Milano, via Ponzio 34/3, I-20133, Milano,Italy}
\address{$^b$ Associazione EURATOM-ENEA, IFP-CNR, Via R. Cozzi 53, 20125 Milano, Italy}
\address{$^c$ ICVBC-CNR, Via Cozzi 54, 20125 Milano, Italy}
%---- Abstract
\begin{abstract}
Pulsed Laser Deposition allows to obtain W and W-Ta alloy coatings with different nanostructures, monitored by X-ray diffraction. The correlation between such structures and the elastic properties is investigated for amorphous-like, ultra-nano- and nano-crystalline coatings obtained by tuning the gas pressure during deposition, annealing temperature and Ta concentration. The full elastic characterization is achieved by surface Brillouin spectroscopy, interpreted by isotropic and anisotropic film models. Amorphous like coatings are obtained with He pressures of tens of Pa. In comparison with bulk W, they have lower stiffness, by about 60\%, closely correlated to the mass density (lower by about 40\%). In the nanocrystalline regime the stiffness is more correlated to the average grain size, approaching the bulk values for increasing crystallite size. Vacuum annealing of amorphous like coatings leads to the nucleation of ultra-nano crystalline seeds, embedded in an amorphous matrix with intermediate values for mass density and stiffness. Here, the stiffness results from an interplay between the crystal size and the density. Alloying with Ta leads to properties which are consistent with the lever rule in the nanocrystalline regime, and deviate from it when the higher Ta concentration, interfering with crystal growth, induces an ultra-nano crystalline structure.

\end{abstract}
%---- Keywords
\end{frontmatter}

\newpage

\section{INTRODUCTION}

\label{Introduction}

Tungsten (W) and W-alloys coatings exhibit unique features such as high melting point, good heat conductivity, high resistance to thermal loads, high density and corrosion resistance. All these properties make W based materials desirable for a wide range of applications. W binary alloys have been found extremely interesting for high power, high voltage, thermal management and radiation shielding applications \cite{Kobayashi, Ibrahim}. In particular, W-Ni alloys are considered among the most promising material for micro and nano-mechanical systems \cite{Borgia_Acta}, while W-Ta coatings have been among the most attractive materials for applications in the aerospace industry \cite{Wadsworth}. The low sputtering yield and the low tritium retention, make pure W films particularly suitable also for selected applications in nuclear fusion facilities, like tokamaks and the future reactor ITER. In this context, W has been chosen as the candidate plasma facing material for the first wall of the ITER plasma chamber, in particular in the divertor region where extreme thermal loads and radiation fields are expected \cite{Kaufmann,Philips,Cambe}. To this purpose, first studies on the applicability of W coatings in nuclear fusion environments were performed depositing thick W layers on structural materials by plasma spray (PS), which allows to cover wide areas, with high thicknesses and limited deposition time and costs \cite{PS}. However, PS limit is represented by the inability to finely control film structure at the atomic scale, which often results in undesired material properties and failure of the coating \cite{PS_2, PS_3}. Alternatively, in order to finely control film structure, morphology and stoichiometry, a variety of physical- (PVD) and chemical- (CVD) vapour deposition techniques is available for W coatings deposition \cite{Ruset_W, Jihong, Perlado, Girault}. 
Pulsed Laser Deposition (PLD) is one of the most versatile PVD techniques, by which both mono- or multi-elemental films such as metals, compounds and carbon based materials, also replicating complicated or unusual target stoichiometry can be deposited \cite{Ngom_1, Tada_1, Tada_2, Fasa_1, Fasa_2, Maaza, Kham_1}. In addition, it offers various possibilities of tailoring the film structure and morphology and of growing films with thicknesses ranging from few nanometers to micrometers \cite{Zani, Uccello, Gondoni}. In particular, it has been proven that properly tuning deposition parameters it is possible to grow W coatings with different morphologies, from compact to porous, and with different structures, from nanocrystalline (nano-W) to amorphous-like (a-W) \cite{Dellasega}. PLD W coatings have been used as proxy of redeposited W in the frame of plasma-wall interaction \cite{Rihanne, Pezzoli}, and for the growth of micrometric-long W-oxide nanowires \cite{Dellasega_Nano}.
\newline
Different morphologies and micro-structures result in different mechanical properties of the deposited films. The mechanical behaviour of coatings is mainly investigated by nanoindentation and by tensile tests \cite{Kulikovsky,Coddet,Hamid,Walczak,Fasaki}; however, it is well known that the application of these two methods to sub-micrometric films is particularly challenging. Alternative characterization techniques have therefore been developed, also due to the increasing demand of non-destructive mapping of the mechanical properties. Surface Brillouin spectroscopy (SBS) has proven to be a useful tool to measure the elastic properties of films \cite{Rioboo, Garcia}, and particularly of sub-micrometric supported films \cite{Beghi_1997, Blachowicz}. SBS relies on the inelastic scattering between the incident laser photon and an acoustic phonon of the material; it thus measures the dispersion relations of surface acoustic waves (SAWs) of the film-substrate system; from them, for known substrate properties, it is possible to derive the elastic moduli of the film. In the case of transparent systems bulk acoustic waves can also be measured, but in the case of metallic films only SAWs are detectable, meaning that in many cases only very few acoustic modes can be measured. The amount of available information is therefore limited, and a complete elastic characterization can require the combination with other techniques \cite{Garcia}.
\newline
However, using suitable substrates, it is possible to confine several modes within the film. Beside their wide availability, Si substrates have acoustic properties which are significantly different from those of W and W-compounds films, mainly because of the huge difference between the mass densities. Due to this difference, W coatings effectively act as multiple mode waveguides, able to confine several acoustic modes. SAWs therefore behave as self-tuning probes, which intrinsically confine themselves predominantly in the layer of interest. The ability to measure several acoustic modes provides a larger amount of information, allowing a more complete elastic characterization.
\newline
In our work we deposited, by PLD, W and W-Ta films on silicon (Si) substrates. Different nanostructures have been obtained exploiting three different variants of film production process: (i) tuning of He pressure in the deposition chamber, (ii) vacuum thermal annealing of a-W, and (iii) alloying with percentages of tantalum, up to 24\%. It turns out that the different nanostructures can be grouped in three different classes: the amorphous-like (a-W), the ultra-nanocrystalline (u-nano-W) and the nanocrystalline (nano-W).  Then, we investigate the relationship between the elastic properties and the nanostructure of these W and W-Ta films. 

%-----

\section{EXPERIMENTAL TECHNIQUES}

\label{Experimental techniques}

\subsection{Film deposition, thermal annealing treatments and structural characterization}

W coatings are produced by PLD, exploiting the fourth harmonic of a Nd:YAG laser with wavelength $\lambda$ = 266 nm, repetition rate of 10 Hz, pulse width of 5 ns and energy per pulse of 150 mJ. The laser with angle of incidence 45 degrees is focused on a 2 in. diameter W target, which is properly rotated and translated in order to ensure a uniform ablation.  The spot area of the impinging laser on the target is about 9 mm$^2$, the laser fluence is 1.6 Jcm$^{-2}$. The species ablated from the W target expand in the deposition chamber where a He background (purity 99.999\%) gas is present. He pressure ranges between 10$^{-3}$ Pa base pressure (vacuum) and 60 Pa. The expanding species, mainly single atoms in the chosen ablation regime, are collected on a substrate, Si(100), positioned 60 mm away from the target. The substrate is aligned with the center of the plume and it is not moved. This setup has been exploited to investigate the effect of different He background gas pressure on the crystalline structure of the growing film. %Varying He pressure from 10$^{-3}$ Pa (vacuum) to 60 Pa, the nanostructure changes from nano-W (vacuum) to u-nano-W (around 5 Pa) and finally to a-W (above 30 Pa). 
\newline
Amorphous-like W coatings have been also deposited using the second harmonic of a Nd:YAG with $\lambda$ = 532 nm, 10 Hz repetition rate, pulse width of 7 ns and, energy per pulse of 815 mJ, laser spot area 8.7 mm$^2$ and laser fluence of 9.3 J cm$^{-2}$. As in the previous case the laser beam impinges on a 2 in. diameter W target.
%A slightly different setup, presented in \cite{Dellasega}, has been exploited to investigate the effect of annealing. First, a-W coatings are deposited with $\lambda$ = 532 nm, pulse width of 7 ns and energy per pulse of 815 mJ. 
He background gas pressure is fixed at 40 Pa and the target to substrate distance is again 60 mm. More details are reported in \cite{Dellasega}. The deposited samples are then annealed in vacuum (10$^{-3}$  Pa) for two hours, at temperatures ranging from 720 K to 1070 K. 
\newline
The effect of Ta alloying has been investigated by the second PLD setup, but with a fixed deposition pressure of 10$^{-3}$ Pa, exploiting a heterogeneus target similar to those described in \cite{Orii}. It consists of a W disk (purity 99.9\%), with a superposed grid obtained from a Ta wire (purity $>$ 99.95\%) of 0.25 mm diameter. The target is rotated and translated, as in the previous case, to ensure uniform ablation. The amount of Ta in the deposited film is varied tuning the pitch of the Ta grid.
\newline
Film thicknesses, determined by cross sections using a Zeiss Supra 40 field emission SEM with an accelerating voltage of 3 - 5 kV, range from 200 nm to 2 $\mu$m depending mainly on deposition time. Coating thickness distribution is Gaussian centered in the middle of the sample where a homogeneous area of about 1 cm$^2$ is present. Elemental composition of W-Ta films has been assessed using Energy Dispersive X-ray Spectroscopy (EDXS). The film structure is analyzed by X-ray Diffraction spectroscopy (XRD) using a Panalytical X'Pert PRO X-ray diffractometer in $\theta $/2$\theta $ configuration. 

\subsection{Mechanical characterization}

\begin{figure}[!t]
\centering
\includegraphics[scale = 0.35]{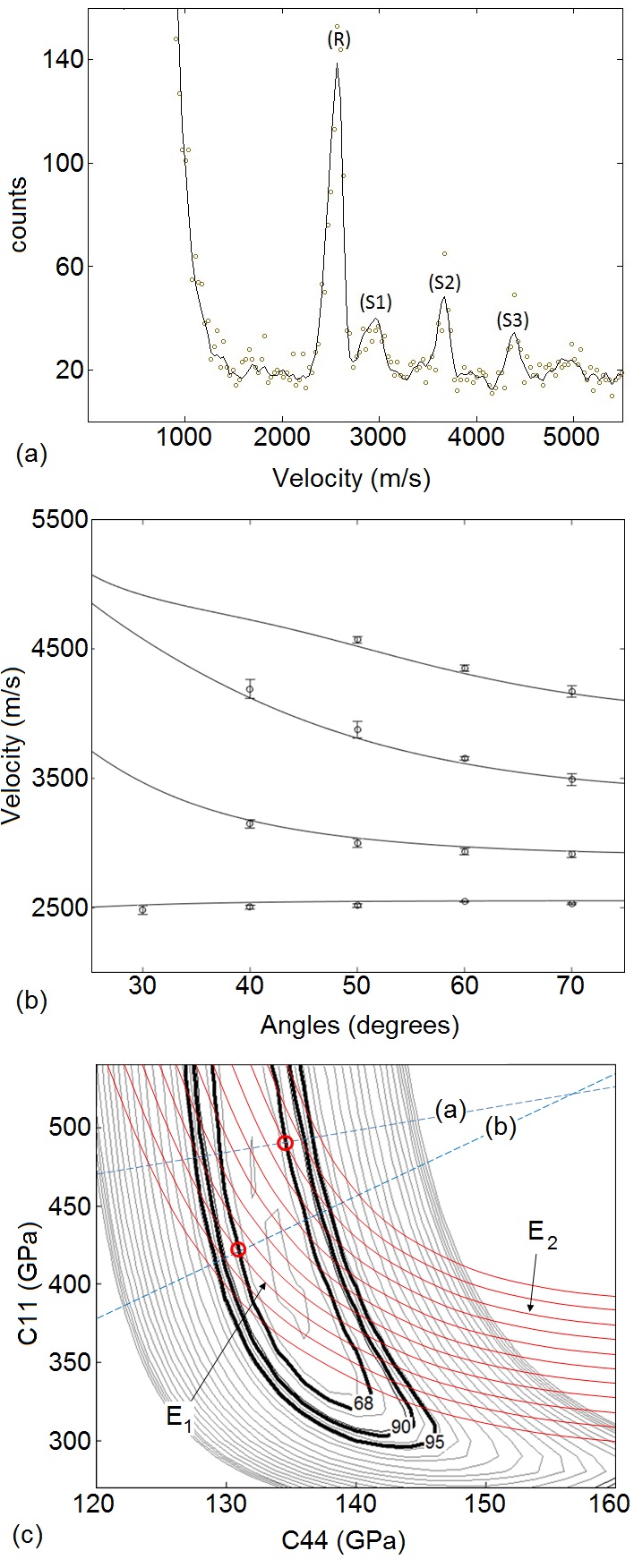}
\caption{a) A typical Brillouin spectrum of a W film on Si(100) substrate. b) Experimental and computed dispersion relation for a W coating. The dots represent the experimental phase velocities while the lines are the theoretical dispersion relations. (c) Isolevel curves for the $LS$ estimator; the curves corresponding to the 68\%, 90\% and 95\% confidence regions are highlated. The lines (a) and (b) refer respectively to (K$_{bulk}$) and ($G/K_{bulk}$).}
\label{Brill}
\end{figure}

SBS is performed with the experimental setup described in more detail in 
\cite{Beghi_2011}. An \textit{Ar$^{+}$} laser is exploited, operating at 200
mW at a wavelength $\lambda _{0}=514.5$ nm. The laser is focused on the central region of the deposited film where the thickness is homogeneous. The focusing spot size is of the order of tens of micrometers. The measurement is thus sensitive to the properties of the film in this area. The backscattering geometry is
adopted, in which the exchanged wavevector $k_{\parallel}$ is determined by
the incidence angle $\vartheta $ as $k_{\parallel}=2\left( 2\pi /\lambda
_{0}\right) \sin \vartheta $. The samples are oriented in order to
investigate propagation along the Si[110] direction, with incident light
polarized in the sagittal plane. Scattered light is collected without
polarization analysis and analyzed by a tandem multipass Fabry-Perot
interferometer of the Sandercock type, typically with a free spectral range
(FSR) of 16 GHz. 
\newline
A typical Brillouin spectrum of a nano-W film on Si(100)
substrate, plotted as a function of the measured velocity $v_{m}=\omega
/k_{\parallel }$, is shown in Fig.\ref{Brill}.a; $\omega $ is the circular
frequency measured by the spectrometer. As noted above, several branches of the
dispersion relation can be measured; the lowest velocity branch is
the Rayleigh wave ($R$), modified by the film itself, and the other branches
are due to Sezawa waves ($S$). The experimental dispersion relations $%
v_{m}(k_{\parallel },j)$, where $j$ is the branch index, are obtained
recording the spectra at different values of $k_{\parallel }$, i.e. at
different incidence angles $\vartheta $ (Fig. \ref{Brill}.b); for each value
of $v_{m}$ an uncertainty $\sigma (k_{\parallel},j)$ can be estimated. 
%%----------------
If the film is modeled as a homogeneous continuum with perfect adhesion to
the substrate, the velocities of the acoustic modes can be computed by the
Christoffel's equations, as a function of the mass densities of the film and
the substrate ($\rho ^{f}$ and $\rho ^{s}$), of the elastic constants of the
film and the substrate ($C_{ij}^{f}$ and $C_{ij}^{s}$), of the film
thickness $t$ and of the exchanged wavevector $k_{\parallel }$: $v_{c}=v_{c}(\rho ^{s},C_{ij}^{s},\rho ^{f},C_{ij}^{f},t|\,k_{\parallel },j)$. 
Once the substrate properties are known, the wavevector is known by the
angle $\vartheta $, the thickness is measured, the computed velocities are
functions of the film properties alone: $v_{c}=v_{c}(\rho
^{f},C_{ij}^{f}|\,k_{\parallel },j)$. In principle, the film properties can
be thus identified by minimizing the least squares estimator: 
\begin{equation}
LS(\rho ^{f},C_{ij}^{f})=\sum_{k_{\parallel },j}\frac{\left( v_{c}(\rho
^{f},C_{ij}^{f},|\,k_{\parallel },j)-v_{m}(k_{\parallel },j)\right) ^{2}}{\sigma ^{2}(k_{\parallel },j)}  \label{estimator}
\end{equation}
\noindent where the sum extends over all the exchanged wavevectors $k_{\parallel}$ and all the measurable
branches.
The derivation of elastic moduli by the minimization of the $LS$ estimator,
discussed in detail elsewhere \cite{Prep}, is outlined here; it is performed
numerically by a MATLAB code developed to this purpose. The number of
independent free parameters depends on the symmetry of the elastic continuum
which represents the film. The code fully accounts for the cubic symmetry of
the Si substrate: the three independent elastic constants are taken as $C_{11}^{s}$,$C_{12}^{s}$ and $C_{44}^{s}$, and are well known. For the film,
the simplest model is the isotropic one, which is considered first. It is
fully characterized by two independent elastic constants, taken as $C_{11}^{f}$ and $C_{44}^{f}$. If the mass density $\rho ^{f}$ is kept fixed,
minimization in the 2-D space $\left( C_{11}^{f},C_{44}^{f}\right) $ returns
the map of the isolevel curves of the estimator $LS$ (Fig. \ref{Brill}.c).
If also $\rho ^{f}$ is left as free parameter, minimization in the 3-D space 
$\left( C_{11}^{f},C_{44}^{f},\rho ^{f}\right) $ returns one map of the type
of Fig. \ref{Brill}.c for each value of $\rho ^{f}$. The minimum of the $LS$
estimator identifies the most probable value of the $\left(
C_{11}^{f},C_{44}^{f}\right) $ couple or the $\left(
C_{11}^{f},C_{44}^{f},\rho ^{f}\right) $ triple. The isolevel curves (or
surfaces) of the estimator, at levels determined by estimation theory \cite{Seber},
supply the confidence regions at any pre-determined confidence level. The
confidence regions turn out to be the most robust result of the
minimization, the minimum being more sensitive to experimental and
discretization errors.
In both the Rayleigh and the Sezawa waves the shear component prevails;
consequently, the confidence region supplies a good estimate (a narrow
interval) for the shear modulus $G=C_{44}^{f}$, and a poor estimate (an
unreasonably wide interval) of the bulk modulus $K$. The indication coming
from the confidence region is therefore supplemented by two physical bounds,
set by the stiffness of bulk crystalline W. It is assumed that the bulk
modulus $K$ cannot exceed the value of crystalline W, and that also the
ratio of the shape stiffness to the volume stiffness, $G/K$, cannot exceed
the value of crystalline W. The ratio $G/K$ is directly related to the
Poisson's ratio $\nu $ \cite{Greaves}: 
\begin{equation}
\nu =\frac{3/2-G/K}{3+G/K}
\end{equation}
\noindent such that the upper bound for $G/K$ is a lower bound for $\nu $. It must be remembered
that the ($G/K$) ratio can be seen as an index of the ductility \cite{Greaves} of the
material.
The 68\% confidence region is considered; its parts which exceed one of the two limits are
discarded, and the remaining part has a quadrilateral shape (Fig. \ref{Brill}.c). The center of this quadrilateral is taken as the best estimate, and the
half amplitudes of the encompassed intervals are taken as estimates of the
uncertainties. This procedure, which is quite obvious for the $\left(
C_{11}^{f},C_{44}^{f}\right) $ couple, is adopted also for any other
modulus, as shown in (Fig. \ref{Brill}.c) for the Young modulus.

\subsubsection{Modelling improvement: anisotropic films}

The estimates obtained by the isotropic model for the films can be improved
by a better consideration of their nanostructure. Nano-W films grown in
vacuum or at very low He pressure tend to show a columnar structure. Since
the columns are narrower than the investigated acoustic wavelengths, the
film modeling by an equivalent homogeneous continuum of average properties \cite{Ramesh}
remains appropriate, but the isotropic continuum is not fully adequate.
Irrespective of the specific crystalline anisotropy of each column, the
columnar films are statistically isotropic in plane, and anisotropic in the
normal direction. The overall symmetry of the film is therefore hexagonal:
this symmetry guarantees in plane isotropy of the elastic properties. 
Full characterization of the elastic properties in the hexagonal symmetry
requires five independent parameters, which can be taken as $\left(
C_{11}^{f},C_{13}^{f},C_{33}^{f},C_{44}^{f},C_{12}^{f}\right) $ and would
need a significant amount of information \cite{Beghi_book}, even though both
the Rayleigh and the Sezawa waves are insensitive to the value of $C_{12}^{f}$.
A full minimization of the $LS$ estimator in a 4-D or 5-D space would be
too expensive, and would require more information than available from the
measured spectra. Therefore a refinement of the results from the isotropic
model has been obtained starting from the values of the $\left(
C_{11}^{f},C_{44}^{f}\right) $ couple obtained by that model (and of $C_{12}^{f}$, which is derived by the isotropy condition), keeping them
fixed, and allowing $C_{13}^{f}$ to $C_{33}^{f}$ to differ from their values
in the isotropic case, which are respectively $C_{12}^{f}$ and $C_{11}^{f}$.
Minimization in the $\left( C_{13}^{f},C_{33}^{f}\right) $ space obviously
allows an improvement of the fit of the dispersion relations. In the
hexagonal symmetry the Young modulus $E$ becomes direction dependent.
Exploiting the functional dependence of $E$ on the $C_{ij}^{f}$ and the
direction \cite{Grimvail} the values of $E$ and $\nu $ \ were derived. The
intervals turned out to be of modest amplitude, and their upper limit,
meaning an upper bound for the stiffness of the homogeneous equivalent
continuum which represents the film, was picked for the following analyses.
The values of the Young modulus $E$, shear modulus $G$ and the ratio of the
shear to bulk moduli ($G/K$) are thus investigated. 

%%-----
\begin{table}[b]
\centering {\relsize{-1.5} 
\begin{tabular}{lcccc}
\toprule & E (GPa) & G (GPa) & K (GPa) & $\nu$ \\ 
\toprule W$_{poly}$ & 404 & 157 & 323.8 & 0.282 \\ 
Measured & 387 $\pm$ 17 & 152.5 $\pm$ 4.5 & 307 $\pm$ 16 & 0.288 $\pm$ 0.001
\\ 
\bottomrule &  &  &  & 
\end{tabular}
}
\caption{Measured elastic moduli of bulk polycrystalline W.}
\label{tab1}
\end{table}

\section{RESULTS AND DISCUSSION}

The procedure we adopted to measure the elastic properties is validated by the analysis of a bulk polycrystalline pure W sample. Since bulk polycrystalline W is elastically globally isotropic, the isotropic symmetry model is appropriate. With the mass density fixed at the bulk value of 19.25 g cm$^{-3}$, we obtained the values of the elastic moduli summarized in Tab. \ref{tab1}, which are in excellent agreement with those proposed in literature \cite{Bernstein}.
%---
\subsection{Background gas pressure: towards the amorphous-like regime}

During the PLD process an adiabatic expansion takes place due to the high pressure and temperature (e.g.  10$^6$ Pa, 5 10$^3$ K)  of the ablated material. The expanding species push away the He gas present in the vacuum chamber. Helium gas becomes increasingly compressed, (\textit{snow-plough effect})  and eventually forms a shock wave \cite{Chrisey}. In this regime interactions between ablated species and background gas molecules are very weak. During the expansion the internal pressure drops down and, once the equilibrium with the background gas is reached, an efficient diffusion and mixing of the ablated species with the background gas takes place. The kinetic energy distribution of the expanding species exhibits a sudden change moving from the shock-wave regime to the diffusion regime. The film properties, morphology and structure, are related to the energy of the impinging species. At high energy the growth of nanocrystalline and compact films is favoured, at low energy the deposited film is porous and exhibits a disordered crystalline structure \cite{Dellasega_App}. In the present case we set PLD process parameters (laser fluence, pressure and target-to-substrate distance) with the aim of obtaining a compact film but with a disordered structure. Since expansion features are related to the atomic mass and the pressure of the background gas (heavy gas is very efficient in quenching kinetic energy) He has been chosen in order to have a fine tuning of the kinetic energy, and consequently structure and morphology. A more detailed discussion of the morphology and structure of coatings obtained with different He pressures is reported in \cite{Dellasega}.  
%---
\begin{figure}[!t]
\centering
\includegraphics[scale = 0.35]{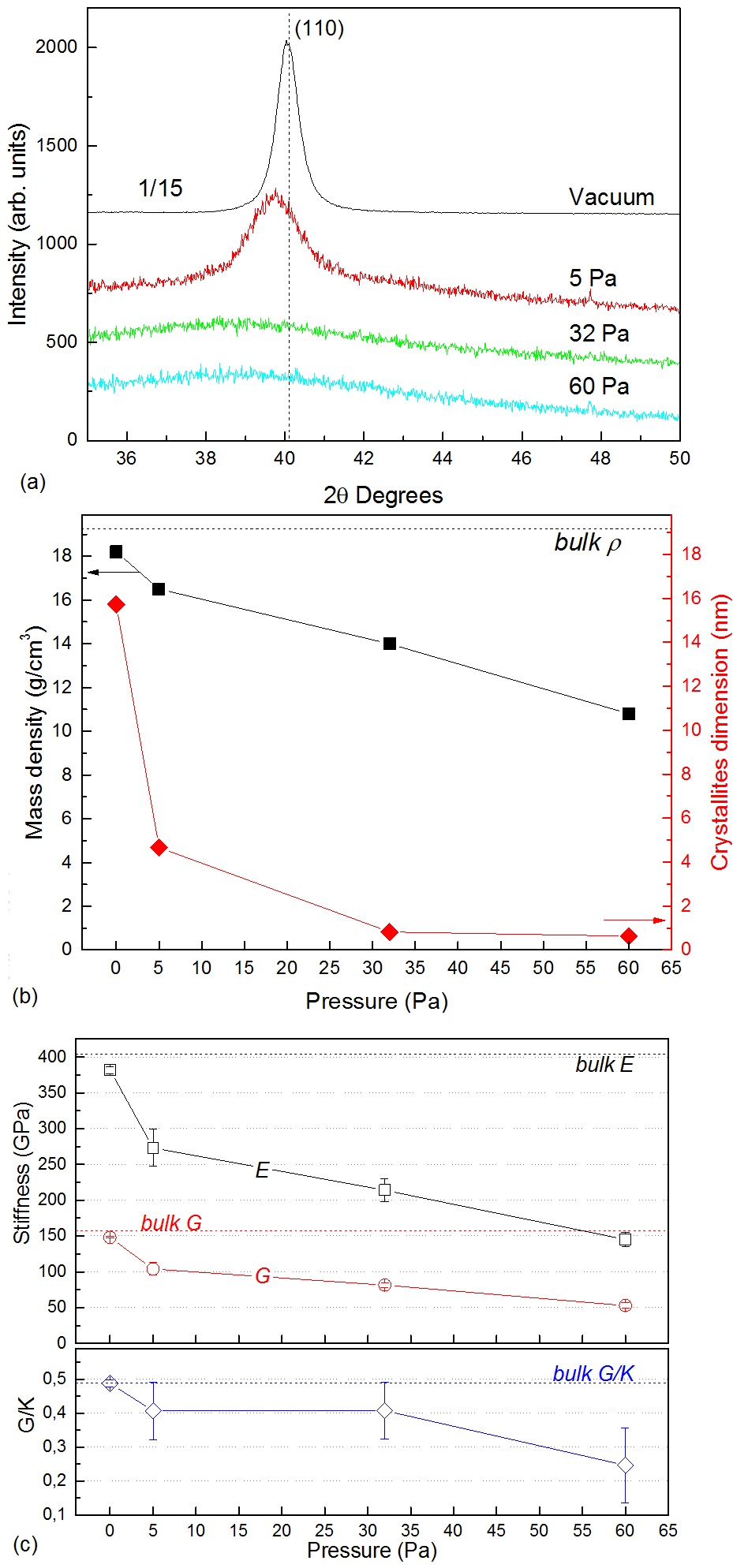}
\caption{Fig.(a): XRD spectra of W nanostructured films; Fig. (b): trends of mass density and crystallites dimension versus He pressure; Fig. (c): trends of E, G and G/K on He pressure.}
\label{11}
\end{figure}
\noindent XRD spectra of the samples deposited with pressures ranging from vacuum to 60 Pa of He are summarized in Fig. \ref{11}a. The spectrum related to the sample deposited in vacuum clearly shows a strong reflection around 40$^\circ$ (2$\theta$). This peak has been assigned to the (110) reflection of $\alpha$-W. The absence of the other reflections (e.g. (200) reflection) is related to the crystallographic oriented growth of the PLD films. The deposition by PLD of metallic films with an oriented crystallographic growth has already been reported in the case of fcc metals like Au \cite{Irissou}, Pt \cite{Riabinina} and Rh \cite{Passoni_M} where the main crystallographic growth direction was the (111). For fcc crystals, the (111) is the direction of minimum configurational energy \cite{Koslowski}. In the case of W, which is a bcc crystal, the atomic planes perpendicular to the (110) direction exhibit the highest planar density, in agreement with the observed direction of growth. The film deposited at 5 Pa exhibits a broader peak, still close to the (110) reflection, but shifted towards lower angles. At 32 Pa and 60 Pa the spectra are different from the previous ones: a broad band appears, the band width being comparable to the ones of amorphous metals. 
\newline
By means of the Scherrer equation we obtain the mean crystallites size $\bar{D}$. In Fig. \ref{11}b the trends of $\bar{D}$ and the film mass density $\rho$ versus He pressure are summarized. The values of $\rho$ were determined by the numerical procedure described above. The trend of mass density is qualitatively in agreement with the results obtained using quartz microbalance \cite{Dellasega}. 
%The establishment of the amorphous regime is marked by the noticeable shrinkage of the crystallites and the decrease of mass density. 
The samples deposited in vacuum are characterized by a mean global density of 18.2 g cm$^{-3}$ that is 5\% lower than the polycrystalline value of 19.25 g cm$^{-3}$ and crystallites size of about 16 nm. This kind of sample, with a crystalline domain size between 10 nm and 16 nm, will be called nanocrystalline W (nano-W). Increasing He pressure to 5 Pa leads to the decrease of the crystallites size from 16 nm to around 4 nm, without significantly affecting mass density (17 g cm$^{-3}$). This growth regime, where crystallite domain size ranges between 4 nm and 10 nm, has been called ultra nano crystalline (u-nano) regime. Higher He pressures result in a further drop of $\rho$ to 14 g cm$^{-3}$ at 32 Pa and to 11 g cm$^{-3}$ at 60 Pa and $\bar{D}$ from 4 nm to approximately 1 nm respectively. Such small values of $\bar{D}$ indicate the presence of at most crystalline seeds; the low values of $\rho$ indicate that the amorphous fraction is predominant. This is related to a higher free volume fraction within the film, meaning a higher mean interatomic distance and a lower mean interatomic binding energy. The last drop of $\rho$ suggests a further decrease of the crystalline fraction, since the crystallites size does not change significantly. This will be called here amorphous-like condition (a-W).
\newline
This evolution of the nanostructure correspond to the variation of the elastic behaviour: trends of the elastic moduli versus He pressure are shown in Fig. \ref{11}c. $E$ drops from 381 to 145 GPa and $G$ similarly drops from 148 to 53 GPa. Accordingly, $G/K$ varies from 0.48 (nano-W) to 0.24 (a-W). Lower stiffness is consistent with the lower values of $\rho$ and $\bar{D}$ proper of the structures discussed above. For the same reasons, also nano-W is characterized by $E$, $G$ and $G/K$ values slightly below the bulk ones (see Tab. \ref{tab1}). 
\newline 
The significant variation of the elastic properties induced by the transition from crystalline to amorphous nanostructure has been widely studied through recent years. Alcala et al. \cite{Alcalà} found a drop in $E$ of about 70\% when going from $\gamma$-Al$_2$O$_3$ to amorphous alumina. Jiang et al \cite{Jiang_BMG} found, for amorphous metals, $G/K$ values 30\% lower than those of the polycrystalline forms. As already mentioned, this decrease of $G/K$ can be interpreted qualitatively as an index of an increased ductility. In the case of amorphous metals, plasticity is attributed to the ability of the material to form shear bands analogous to the slip systems of their crystalline counterparts. The short-range atomic order of amorphous metals can determine the elastic moduli in a way different from that of the crystalline phase, where a long range order is also present \cite{Wang_Acta}. In this way amorphous metals can be macroscopically brittle but microscopically capable of sustaining plastic shear flow \cite{Schuh_Acta}; in these terms, the observed decrease of $G/K$ can be interpreted as an index of increased ductility with the growth of the amorphous phase. Our results are in qualitative agreement with the observed decrease of the $G/K$ ratio going from nanocrystalline to amorphous phases \cite{Alcalà, Jiang_BMG}.

%%%%%%%%%%%%%%%%%%%%%%%%%%%%%%%%%%%%%%%%%%%%%%%%%%%%%%%%%%%%%%%%%%%%%%%

\subsection{Thermal annealing: effects of crystallization}

Using a slightly different laser setup (see Section 2.1) we produced a-W film with a background He pressure of 40 Pa. The higher threshold pressure for the formation of the amorphous-like phase, compared with the previous case, is related to the higher laser fluence used in this setup. The deposited samples exhibit a crystallite size of less than 2 nm and a mass density below 12 g cm$^{-3}$, indicating the presence of a significant void fraction within the films. More details are present in \cite{Dellasega}. Samples are then subjected to thermal annealing; 
%%%%%%%%%%%%%%%%%%%%%
\begin{figure}[!t]
\centering
\includegraphics[scale = 0.35]{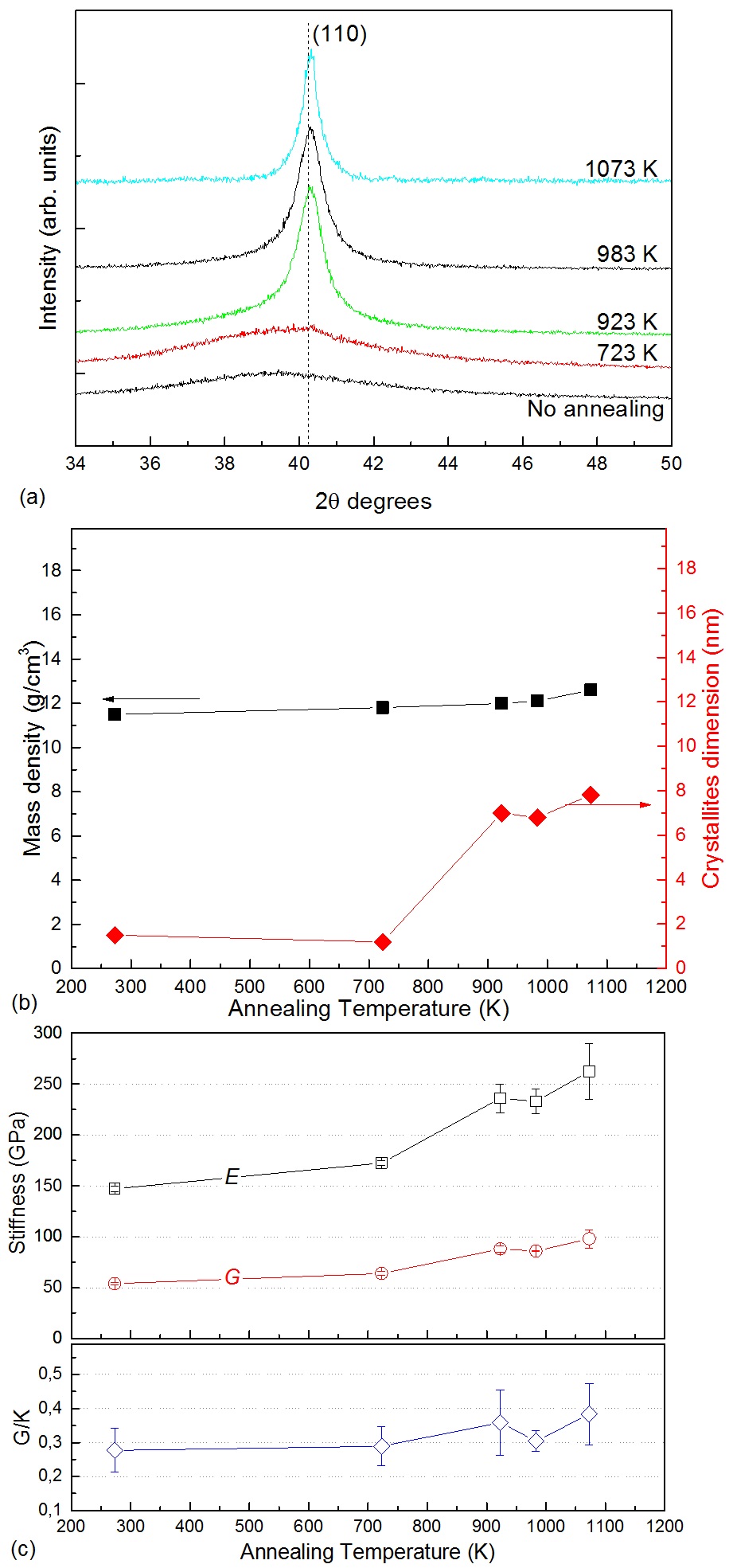}
\caption{Fig.(a): XRD spectra of $a-W$ films exposed to annealing processes; Fig. (b):  trends of mass density and crystallites dimension versus annealing temperature; Fig. (c): trends of E, G and G/K on annealing temperature.}
\label{12}
\end{figure}
%%%%%%%%%%%%%%%%%%%
\noindent it is well known that this type of treatment promotes crystallization which, in turn, can be related to a progressive stiffening and embrittlement \cite{Kuman_Acta}. This can be attributed to different phenomena such as structural relaxation, free volume annihilation, phase separation and short-range ordering. Annealing kinetics depends on temperature: below the glass transition temperature T$_g$ the process is governed by structural relaxation, while for T $>$ T$_g$ crystallization prevails. Furthermore, if thermal annealing is performed above the recrystallization temperature treshold, it can lead to the complete recovery of bulk properties \cite{Sung, Chia, Zhang_2008}. 
\newline
The effect of thermal annealing temperature is investigated performing the treatment in vacuum for 2 hours, at temperatures between 720 K and 1070 K. 
\newline
XRD spectra are shown in Fig. \ref{12}a. At 720 K the spectrum remains almost identical to that of the as-dep a-W, except for a small peak appearing in correspondence of the $\alpha$-W (110) reflection. This feature indicates that the defects diffusion is already active at a temperature below one half of the recrystallization temperature of bulk W ($\sim$1700 K), although it is not yet able to radically change the structure of the film. Instead, for annealing temperatures at or above 920 K the amorphous band essentially desappears in favour of the peak associated to the crystalline $\alpha$-W phase, which becomes better defined. The mean crystallites dimension, reported in Fig.\ref{12}b, remains below 2 nm till 720 K and then suddenly changes over 920 K, remaining approximately constant around 8 nm until 1070 K. This shows that, at or slightly below 920 K, a-W films undergo partial crystallization and the coatings become u-nano-W, D being about 8 nm. However, the mass density, determined by SBS numerical data analysis,  undergoes only a limited variation, going from the as-dep value of 11 g cm$^{-3}$ to 12.8 g cm$^{-3}$ at 1070 K (Fig. \ref{12}b), as confirmed by the fact that upon annealing, no appreciable film thickness variation was observed. Since the nanocrystals have higher mass density than the pristine a-W, this small variation of $\rho$ suggests that nanocrystals remain embedded in more amorphous matrix which retains a significant average void fraction. In parallel to the nucleation of crystalline seeds, the void fraction of the amorphous phase might coalesce, nucleating ultra-nano voids, which however remain too small to be detected by SEM. This small density increase is consistent with the annealing temperature still well below the recrystallization temperature of bulk W, not yet allowing phenomena like e.g. a massive flow of vacancies to the outer surface. The features of such nanostructure could be further elucidated by Grazing-Incidence Small-Angle X-ray Scattering, which clearly distinguishes between the regions with different electron density, or by Positron Annihilation Lifetime Spectroscopy, sensitive to the void dimension and distribution.
\newline
The elastic moduli are also affected by the annealing treatment. An increase in $E$, $G$ and $G/K$ is however expected for temperatures either above or below T$_g$. A trend similar to the one of $\bar{D}$ is observed: at low annealing temperatures the elastic moduli are not significantly modified, while at 920 K, $E$ and $G$ increase by about 60\% ($E$ goes from around 150 GPa to around 230 GPa), with a corresponding rise of $G/K$ by 33\% ($G/K$ from 0.27 to 0.36) as shown in Fig. \ref{12}c. Material stiffening is thus related to the beginning of the crystallization. 
\newline
Finally, it can be seen that the a-W sample annealed at 1070 K (u-nano-W) is characterized by $E$ = 262 GPa, $G$ = 98 GPa and $G/K$ = 0.38. These values remain well below the bulk values. This is consistent with the lower mass density: the stiffer nanocrystals remain embedded in a softer amorphous matrix. A complete recovery, till the microcrystalline state, would require higher annealing temperatures and/or longer annealing times.

\subsection{Ta-W films: alloying effects}

\begin{figure}[!t]
\centering
\includegraphics[scale = 0.35]{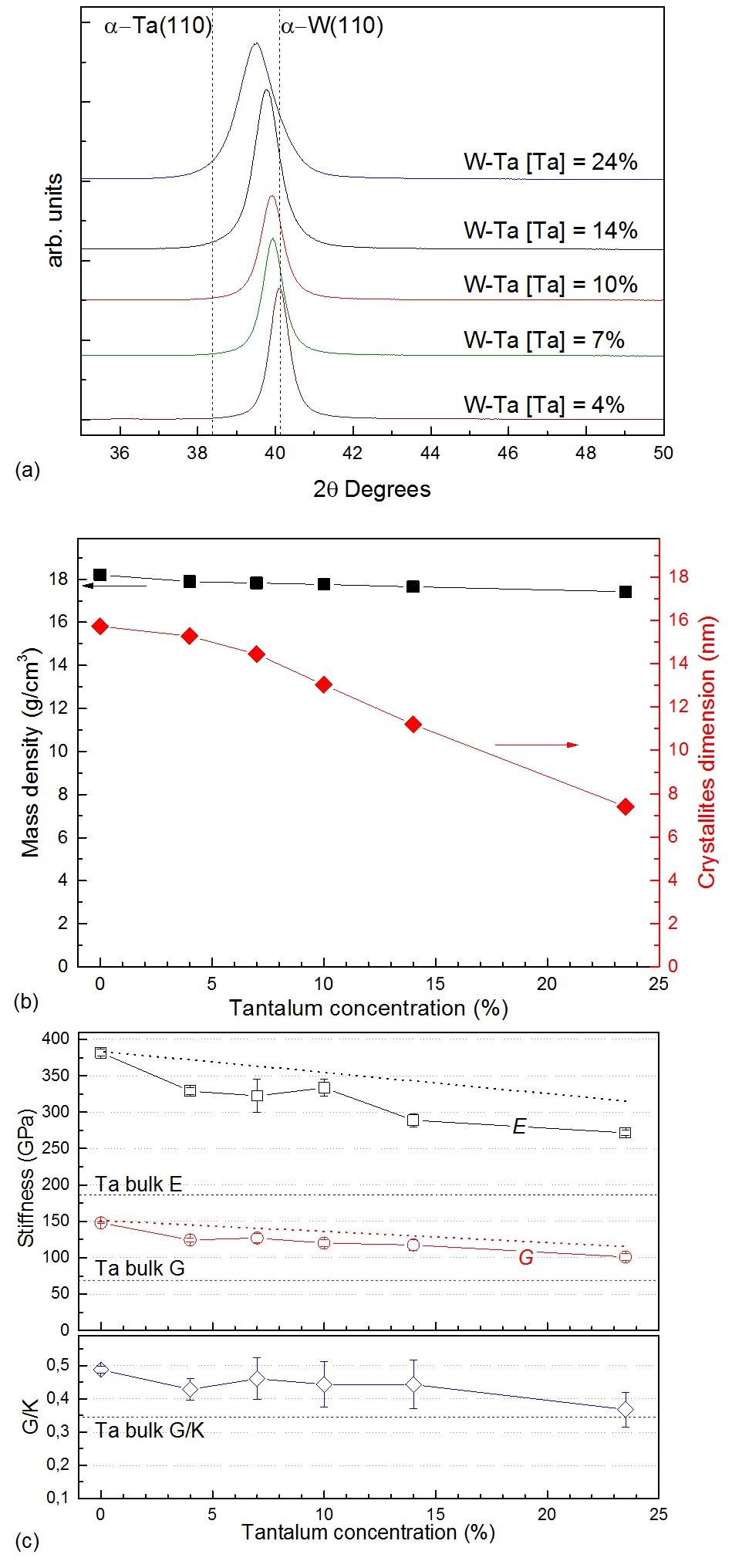}
\caption{Fig. (a): XRD spectra of W-Ta films; Fig. (b): trends of mass density and crystallites dimension versus Ta concentration; Fig. (c): trends of E, G and G/K vs. Ta concentration (dotted lines: values computed bythe lever rule of eq. \ref{eq2})}
\label{13}
\end{figure}

W-Ta films were deposited in vacuum with Ta concentrations ranging from 4 to 24\%.  In the XRD spectra in Fig. \ref{13}a it is evident that when Ta concentration increases no new reflections appear and the main peak, corresponding to the (110) reflection of crystalline $\alpha$-W, gradually shifts from 40.416 - 2$\theta$ degrees, the angle corresponding to $\alpha$-W, towards 38.473 - 2$\theta$ degrees, the angle corresponding to $\alpha$-Ta. As already reported in \cite{Bhattarai, Muzyk, Young}, this shows that the addition of Ta to a W matrix leads to solid solution regime in which the Ta atoms are substitutional. However, boosting the Ta amount leads to a decrease in the crystallinity of the film. Ta atoms interfere with the W lattice hindering the growth of the crystallites. As a result, $\bar{D}$ goes from 15 to 7 nm when the Ta content goes from zero to 24\%. This variation is clearly more appreciable than the mass density one. In this case, since we are dealing with solid solution regimes, the mass density can be computed by the lever rule:
\begin{equation}
\rho_{W-Ta} = [W]\rho_W + [Ta]\rho_{Ta}
\label{eq2}
\end{equation}
\noindent This leads to values of $\rho$ ranging from 18.2 g cm$^{-3}$ (pure nano-W) to 17.4 g cm$^{-3}$ (24\% Ta). 
\newline
No appreciable ductilization of microcrystalline W has been reported upon alloying with Ta \cite{Reith}. We observe trends of $E$ and $G$ which are almost linearly decreasing with increasing Ta concentration (Fig. \ref{13}c). In particular, the material is softened by about 28\%: $E$ falls from 381 GPa to 271 GPa and similarly $G$ from 148 GPa to 101 GPa. The most probable value of the $G/K$ ratio does not substantially change between zero and 14\%, while decreasing by 30\% in the alloy with 24\% of Ta. 
%%%%%%%%%%%%%%%
%%%%%%%%%%%%%%%
Fig. \ref{13}c shows that the gap between the experimental values of $E$, $G$ and $G/K$ and those computed by the lever rule becomes remarkable for Ta concentrations above 10\%. This discrepancy can be attributed to the above mentioned reduction of the crystallites size which leads to a variation of the mechanical behaviour beyond the one foreseen by the solid solution regime alone when the concentration of Ta atoms becomes relevant. Accordingly to the above discussion (see Section 3.1), the decrease of the $G/K$ ratio can be interpreted as indicating the increasing material ability to support local plastic flow. When the Ta fraction increases, the higher elastic moduli of W are gradually shifted towards the lower values proper of Ta, leading to a global softening and local ductilization of the material.
%%%%%%%%%%%%%%%

\subsection{Crystallites dimension and mass density effect}
\begin{figure}[!t]
\centering
\includegraphics[scale = 0.35]{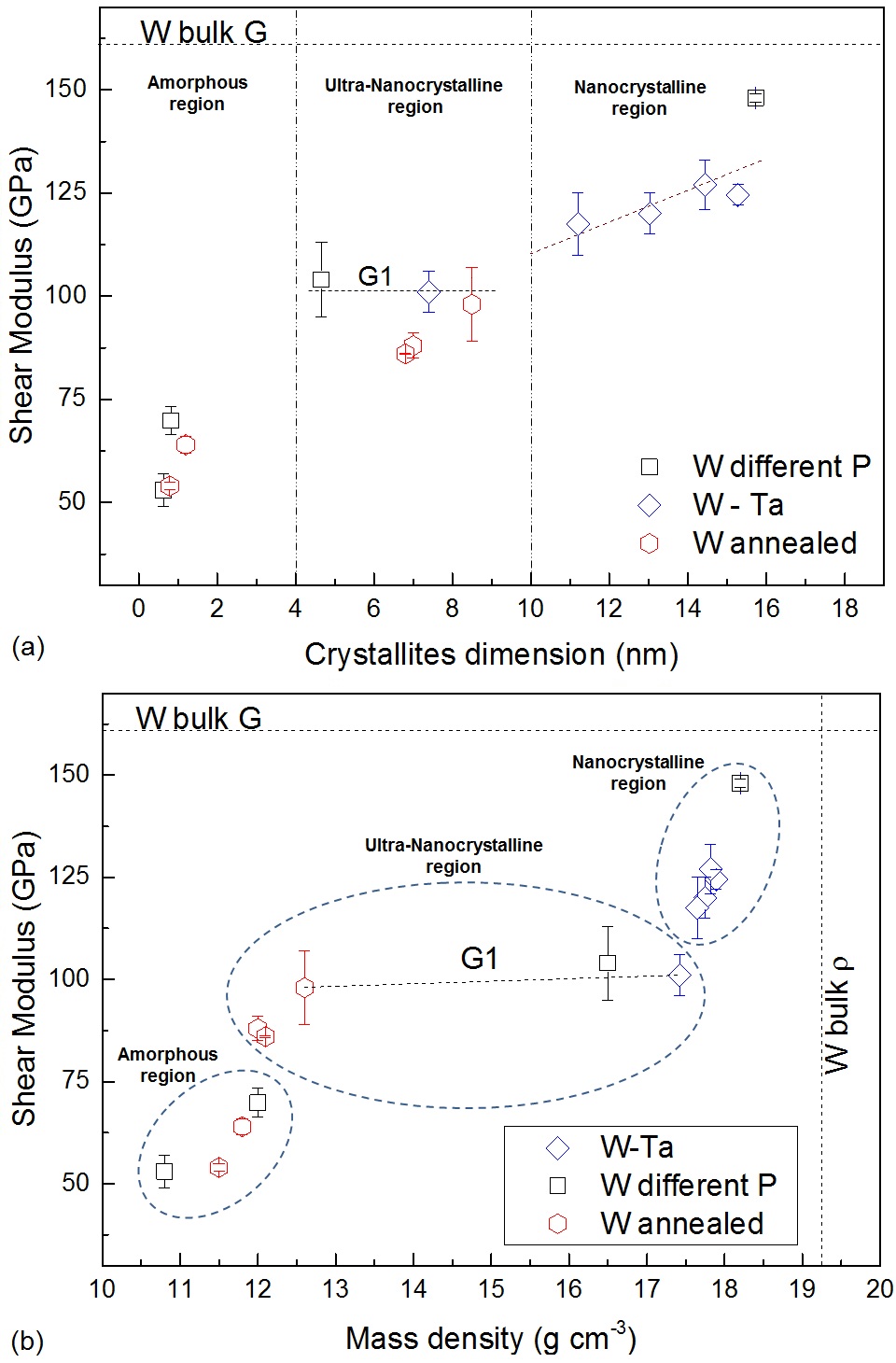}
\caption{Fig. (a):  trends of shear modulus versus mean crystallites size; Fig. (b): trends of shear modulus versus mass density.}
\label{14}
\end{figure}
We have shown above that the elastic moduli of a film can be tuned varying the mass density and the crystallites dimension, simultaneously or independently. This is best shown by the trend of shear modulus $G$ which, as noted above, is the most reliable outcome of Brillouin spectroscopy. Fig. \ref{14} summarizes the trends of $G$ of the all analyzed samples versus $\bar{D}$ (\ref{14}.a) and $\rho$ (\ref{14}.b). 
\newline
The amorphous region ($\bar{D}$ $<$ 4 nm) is characterized by markedly lower values of $G$. In this region the samples behave as metals in the amorphous regime, and the shear stiffness changes appreciably even for a very small variation of $\bar{D}$. The transition to the ultra-nanocrystalline region is characterized by crystallites size becoming of at least few nanometers ($\sim$ 4 nm). This crystallites remain embedded in an amorphous matrix of different elastic properties; the overall shear stiffness is thus determined by both the properties of the crystallites and those of the matrix. This explains the small sensitivity of $G$ to the crystallite size. For $\bar{D}$ above $\sim$10 nm, the nanocrystalline domain is achieved, in which $G$ is almost linear with the crystallites dimension. Extrapolating this linear trend, the mean crystallites size necessary to confer the bulk elastic properties to nano-W turns out to be $\sim$25 nm.
\newline
From Fig. \ref{14}b, it is now evident that in the amorphous region the different stiffnesses are mainly related to the different mass densities. In the ultra-nanocrystalline region, we find a low sensitivity of $G$ to the mass density, the crystallite size becoming also relevant. This region includes samples obtained by the three different preparation processes described above (line $G1$). For this reason their mass density and crystallite size are not correlated, and go instead in opposite directions. The interplay between crystallites and matrix properties is such that the overall stiffness remains, fortuitously, almost constant, for mass densities going from 12.6 to 16.5 g cm$^{-3}$, while the crystallite sizes go from 8 to 4 nm. For the samples analyzed in this study, the growth of the crystallites balances the decrease of $\rho$. Finally, in the nanocrystalline region $G$ turns out to be linearly increasing with $\rho$ as confirmed by the lever rule.

%----
\section{CONCLUSIONS}

Thanks to the PLD versatility we were able to grow W based films with different structures, namely nano-W, u-nano-W and a-W. In particular, tuning the He background gas pressure during deposition, we obtained W coatings from nano-W (in vacuum) to a-W (at 40-60 Pa). Post deposition thermal annealing of a-W samples promotes defects diffusion and crystallites growth, leading to the u-nano-W region. Finally, nano-W alligation with Ta atoms resulted in a solid solution regime where the crystallites size gradually decreases with increasing Ta concentration. These structures have been characterized by XRD analysis: a-W samples have $\bar{D}$ $<$ 4 nm, for $\bar{D}$ between 4 and 10 nm the coatings are u-nano-W, while for $\bar{D}$ $>$ 10 nm the nano-W region is achieved. Beside this wide range of crystallites size values, also the film mass density is deeply affected by nanostructuring: values closed to bulk polycrystalline W (19.25 g cm$^{-3}$) are found for nano-W down to below 12 g cm$^{-3}$ for a-W. In the u-nano-W region, nanometric crystalline seeds are still embedded in a more amorphous matrix which contains a significant void fraction, as indicated by the low mass density. The nature of this void fraction (e.g. disordered amorphous nanostructure, or dispersion of nano-voids) could be elucidated only by a systematic TEM analysis. 
\newline
The different nanostructures determine different elastic properties; SBS has proven to be sensitive to these differences allowing a quantitative characterization of all the elastic moduli and their correlation with the peculiar film structure. The a-W region is characterized by low stiffness, about 40\% of the bulk crystalline values; the differences among various samples are mainly correlated to differences in the film mass density (which is however low), the crystalline order being negligible. For u-nano-W samples, the elastic properties result from a non obvious interplay between the crystalline fraction, the averaged size of the crystalline seeds and void fraction. In this region, even if appreciably different mass densities are found, the stiffnesses are quite similar (around 60\% of bulk W). Finally, the nano-W form is quite compact, as witnessed by the mass density which approaches the bulk value, and the stiffness is mainly determined by the crystallites size. This is comfirmed by the stiffness of W-Ta alloys which is consistent with the lever rule for sufficiently large grains, and deviates from it for smaller grains (u-nano-W).
\newline
The PLD versatility in tailoring the properties of the films is thus comfirmed.

\section*{Acknowledgments}
The authors would like to thank Andrea Giuliani and Tommaso Agosti for part of the Brillouin spectroscopy measurements and Simone Finizio for W-Ta films deposition. This work has received funding from the European Union’s Horizon 2020 research and innovation programme under grant agreement number 633053. The views and opinions expressed herein do not necessarily reflect those of the European Commission. The research leading to these results has also received funding from the European Research Council Consolidator Grant ENSURE (ERC-2014-CoG No. 647554).

\bibliographystyle{model1a-num-names}
\bibliography{<your-bib-database>}

\end{document}